# PROCEEDING BOOK

(Abstracts and Full-Text Papers)

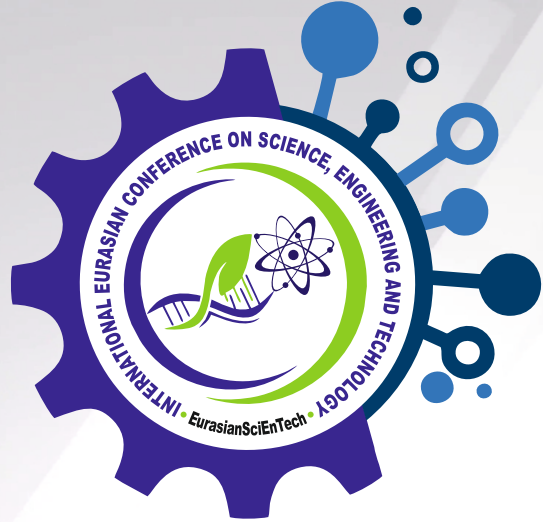

## 4th INTERNATIONAL EURASIAN CONFERENCE ON SCIENCE, ENGINEERING and TECHNOLOGY

(EurasianSciEnTech 2022)

14 – 16 December 2022
Ankara / Turkey

ISBN 978-605-72134-1-9

**EurasianSciEnTech 2022**

www.eurasianscientech.org

➢ **ORAL PRESENTATION**

**Data Augmentation with GAN increases the Performance of Arrhythmia Classification for an Unbalanced Dataset**

Okan Düzyel[1*] (*ORCID: https://orcid.org/0000-0002-9123-3146*),
Mehmet Kuntalp[1] (*ORCID: https://orcid.org/0000-0002-3381-9026.*),

[1]Dokuz Eylül University, The Graduate School of Natural and Applied Sciences, Izmir, Turkey.

*Corresponding author e-mail: okanduzyel@gmail.com

**Abstract**
Due to the data shortage problem, which is one of the major problems in the field of machine learning, the accuracy level of many applications remains well below the expected. It prevents researchers from producing new artificial intelligence-based systems with the available data. This problem can be solved by generating new synthetic data with augmentation methods. In this study, new ECG signals are produced using MIT-BIH Arrhythmia Database by using Generative Adversarial Neural Networks (GAN), which is a modern data augmentation method. These generated data are used for training a machine learning system and real ECG data for testing it. The obtained results show that this way the performance of the machine learning system is increased.

**INTRODUCTION**
In today's world, it has become easy to access data thanks to the development of advanced computers, the spread of the internet and increase in information sharing, However, data in some fields, such as medicine and biomedicine, are insufficient in number due to the lack of data sharing and the purpose of protecting personal data. Re-collecting and acquiring data via sensors is both time-consuming and costly. For such works, electronic devices should be purchased, working systems should be established and necessary permits should be obtained according to the policies. A machine learning model trained with a dataset with insufficient data and unequal sample numbers (i.e. unbalanced data) will not be able to interpret real-life data properly and will give erroneous results. This problem is one of the biggest obstacles to developing artificial intelligence-based devices. To overcome this problem, classical data augmentation methods (Hoelzemann et al., 2021) can be used. However, with these methods, the existing data is changed, and new data is not suitably produced. Although this solution works partially, it does not give us successful results when compared to real-life data. With innovative approaches such as Variational Auto Encoder (VAE) (Kingma & Welling, 2013) and Generative Adversarial Neural Networks (GAN) (Goodfellow et al., 2014), new data can be effectively synthesized. These data may have more resemblance to reality as if they were collected from real life.

With the introduction of VAE, there has been a minor revolution in data augmentation (Saldanha et al., 2022). VAE has a structure based on Auto Encoder (AE). The auto encoder compresses the input data and then tries to reconstruct it. In the meantime, it causes some data loss, but several features are formed on the middle layer, i.e. the latent space, which has the lowest number of neurons. These features may be input-related features, or they may be the information of unwanted noise at the input. For this reason, the AE features can be used for both feature extraction purposes as well as for identifying and blocking noise (Lee et al., 2021). However, there is some data loss as the part where VAE is insufficient compresses the data towards the latent space. This is reflected as blurring in images and muffled sound in audio files (Hou et al., 2017; Zhu et al., 2021)

With the announcement of the GAN, there has been a real revolution in this field. Because GAN, contrary to VAE, does not experience data loss and could synthesize much more realistic data. The working logic of GAN is based on the principle that two deep neural networks compete with each other, i.e. Generator and Discriminator networks. The Generator generates random data and sends the data it generates to the Discriminator. This network decides how similar the generated data is to the real data. If it finds the data to be unrealistic, it trains the generator's neurons by the loss value between the real data and unreal data. As a result of this training, the data produced by the generator starts to be more realistic each time. The loss values of the Generator and Discriminator intersect at a point. This point is called the Nash equilibrium, and at this point the generator has confused the discriminator. At this moment, new data production is finished. GAN was first used in image synthesis (Goodfellow et al., 2014; Ouyang et al., 2018; Siarohin et al., 2017) and later adapted to many fields. With GAN, audio files can be synthesized (Donahue et al., 2018), as well as tabular data (Xu et al., 2019).

Many studies have been done in the literature on ECG signal synthesis with classical methods (Mukhopadhyay & Sircar, 1995; Zigel et al., n.d.). In addition to different methods, there are studies in which GAN based ECG synthesis (Delaney et al., 2019) and VAE based synthesis (Kuznetsov et al., 2020) are performed.

In this paper, ECG data from MIT-BIH Arrhythmia Database (Moody & Mark, 2001), is first generated by a GAN. Then the mathematical analysis of the synthesized data and its classification success are examined.

**MATERIALS AND METHODS**

The MIT-BIH Arrhythmia Database comes with two parts, which are train and test sides, and consists of five classes in total. Due to the unequal distribution of data in different classes, it is a highly unbalanced dataset. The names and sample distributions of train data (light blue) and test data (orange) are shown in Figure1 below.

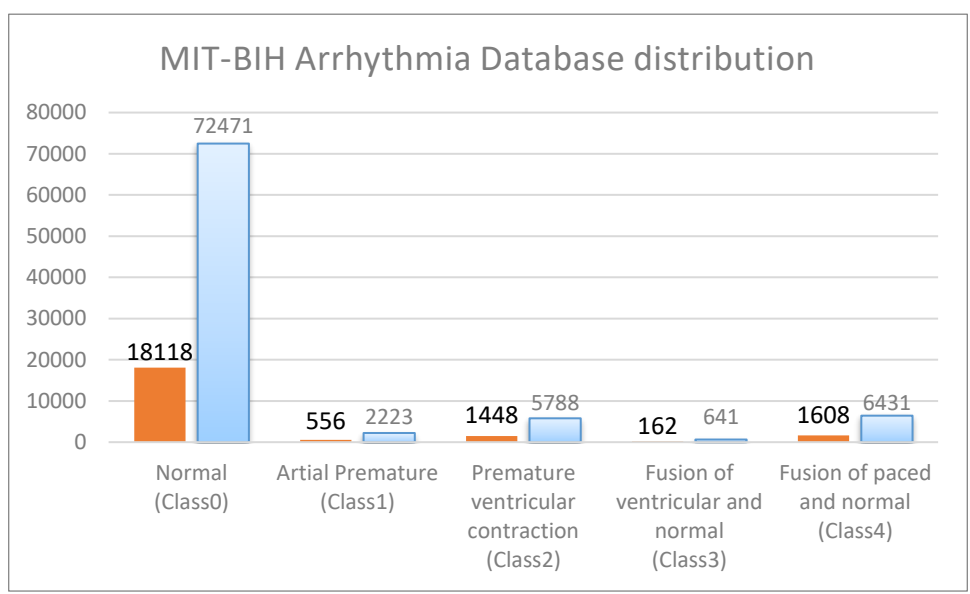

**Figure 1.** MIT-BIH Arrhythmia database distribution

When Figure 1 is examined carefully, the Normal (Class0) class dominates all other classes. This is a typical example of unbalanced data distribution, which is not good for machine learning models. Because the model tends to memorize data in Class 0. This causes it to be unable to successfully separate other classes from this class. A solution is to increase the number of classes that are less in number with classical data augmentation methods. These classical methods used to generate signals can be Time-shifting, Pitch scaling, Noise addition, Low/High/Band-pass filters, and Polarity inversion, but ultimately this approach does not yield successful results for every application.

We designed a GAN to augment the available ECG signals. All ECG signals in the MIT-BIH Arrhythmia Database have been digitized by taking 187 samples, and each sample is correlated with the previous sample value. This shows us that we should consider all the values in the signal and points out that we should use a memory-based model. Therefore, Long Short-Term Memory GAN (LSTM-GAN) can be used to synthesize new ECG signals. Due to its structure, LSTM does not forget previous information and tries to correlate the next values with the previous values. In this way, information such as the rise and fall of the signal, amplitude values, frequency changes are processed. The architecture of the LSTM-GAN model used in this study is shown in Figure 2 below.

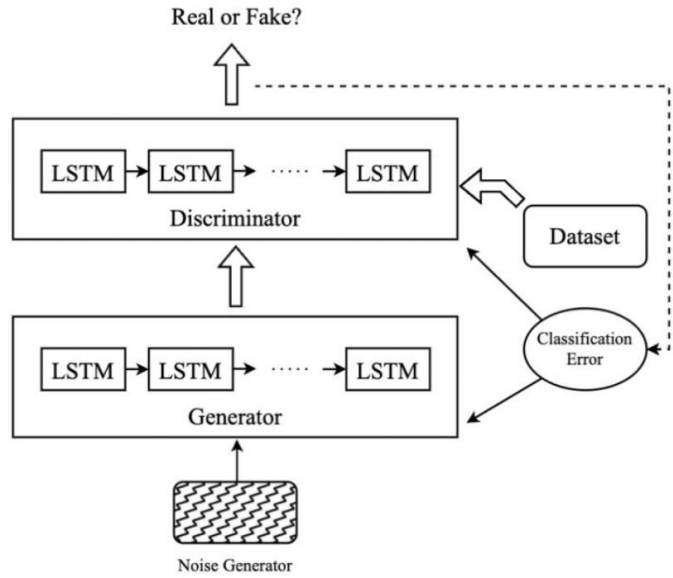

**Figure 2.** LSTM-GAN Architecture

Unlike other machine learning models of GAN, there are two different loss functions defined for generator and discriminator, and these functions are used coherently.

$$L_D = Error(D(x), 1) + Error(D(G(z)), 0) \quad (1)$$

$$L_G = Error(D(G(z)), 1) \quad (2)$$

$$H(p, q) = \mathbb{E}_{x \sim p(x)}[-log q(x)] \quad (3)$$

$$\min_G \max_D V(D, G) = \mathbb{E}_{x \sim p_{data}(x)}[log D(x)] + \mathbb{E}_{z \sim p_z(z)}[\log (1 - D(g(z)))] \quad (4)$$

The difference of the LSTM-GAN model from the classical GAN is that there are LSTM modules inside the Generator and Discriminator units. Apart from that, LSTM-GAN tries to decrease the loss of the Generator, increase the loss of the Discriminator and find the Nash equilibrium value between the two, just like the classical GAN. When this equation is established, new ECG signals are synthesized. Since Class0 is the largest class, there is no need to synthesize new data for this class. Moreover, the make the dataset more balanced, we eliminate uniformly some data from Class0. The following Table 1 shows the number and types of data created by GAN for all classes and final shape of train dataset in this study.

Table 1. Data Augmentation and Balancing for all classes

| Class Name | Original number of train data | Added-Sub. Train Data | Final Number of Train Data | Used Method |
|---|---|---|---|---|
| Class 0 | 72471 | -62471 | 10000 | Down Sampling |
| Class 1 | 2223 | +7777 | 10000 | LSTM-GAN |
| Class 2 | 5788 | +4212 | 10000 | LSTM-GAN |
| Class 3 | 641 | +9359 | 10000 | LSTM-GAN |
| Class 4 | 6431 | +3569 | 10000 | LSTM-GAN |

Now that the train dataset is balanced, it can now be used to train the machine learning model, which is used as the classifier. Since the ECG is a 1D time-domain signal, the 1D Convolutional Neural Network (1D CNN) model is preferred as the classifier. The architecture of the 1D CNN model used in this study is shown below in Figure 3.

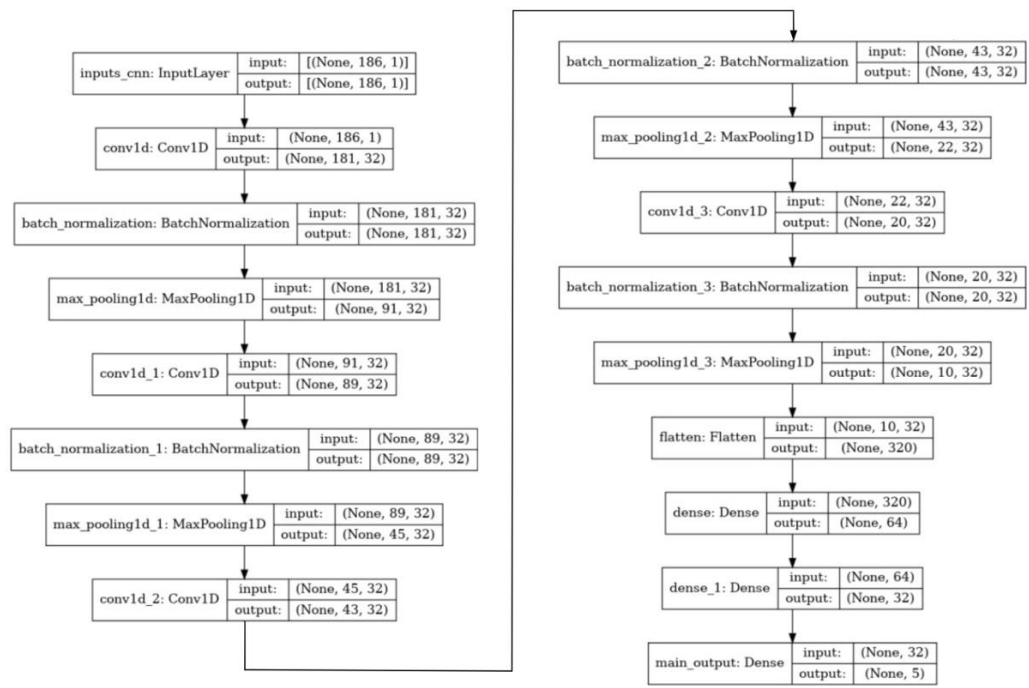

Figure 3. 1D CNN model architecture

**RESULTS and DISCUSSION**

In order to produce new data with LSTM-GAN, each class is augmented separately by a different GAN, each of which is optimized for generating new data especially for one class. Example of the result of these processes is given in Figure 4.

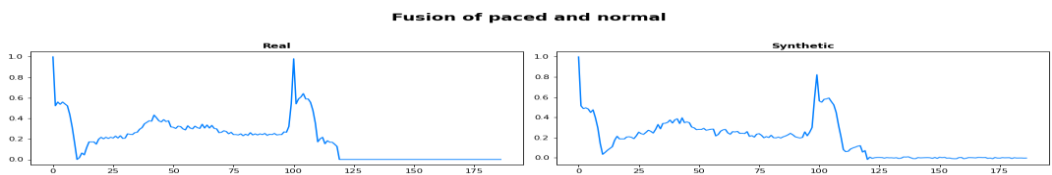

**Figure 4.** A sample of Real and Synthetic ECG output for Fusion of paced and normal class

Batch Size:64 and Epoch:10 was set in the experiments and the test data was not touched in both experiments. This is to make the test data distribution as if it was taken from real life. Number of test data used in classes is shown in Tables 2 and Table 3, which show the experimental results of using original unbalanced and synthetically balanced datasets.

**Table 2.** Classification results of original and unbalanced MIT-BIH Arrhythmia Dataset

|  | MCC | Precision | Recall | F1-Score | Number of Test Data |
|---|---|---|---|---|---|
| **Class 0** | 0.42 | 0.34 | 0.99 | 0.51 | 18118 |
| **Class 1** | 0.56 | 0.93 | 0.44 | 0.59 | 556 |
| **Class 2** | 0.70 | 0.94 | 0.63 | 0.75 | 1448 |
| **Class 3** | 0.29 | 1.00 | 0.11 | 0.20 | 162 |
| **Class 4** | 0.86 | 0.95 | 0.86 | 0.90 | 1608 |
| **Accuracy** |  |  |  | 0.93 | 21892 |
| **Macro Avg** |  | 0.83 | 0.61 | 0.59 | 21892 |
| **Weighted Avg** |  | 0.45 | 0.93 | 0.56 | 21892 |

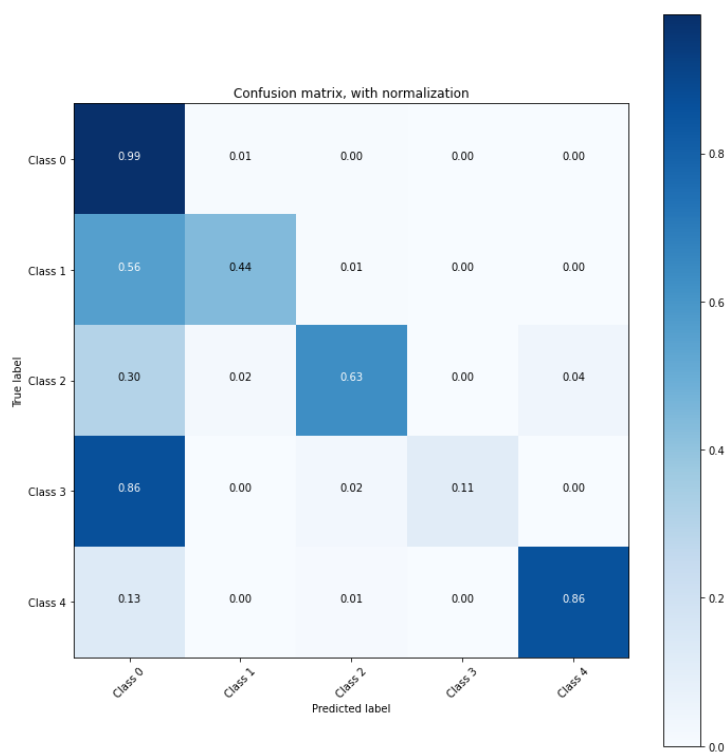

**Figure 5.** Confusion matrix of original and unbalanced MIT-BIH Arrhythmia Dataset

**Table 3.** Classification results of augmented and balanced MIT-BIH Arrhythmia Dataset

|  | MCC | Precision | Recall | F1-Score | Number of Test Data |
|---|---|---|---|---|---|
| **Class 0** | 0.66 | 0.67 | 0.80 | 0.73 | 18118 |
| **Class 1** | 0.75 | 0.88 | 0.72 | 0.79 | 556 |
| **Class 2** | 0.80 | 0.89 | 0.82 | 0.84 | 1448 |
| **Class 3** | 0.79 | 0.79 | 0.91 | 0.84 | 162 |
| **Class 4** | 0.93 | 0.97 | 0.92 | 0.94 | 1608 |
| **Accuracy** | - | - | - | 0.83 | 21892 |
| **Macro Avg** | - | 0.84 | 0.83 | 0.83 | 21892 |
| **Weighted Avg** | - | 0.71 | 0.81 | 0.75 | 21892 |

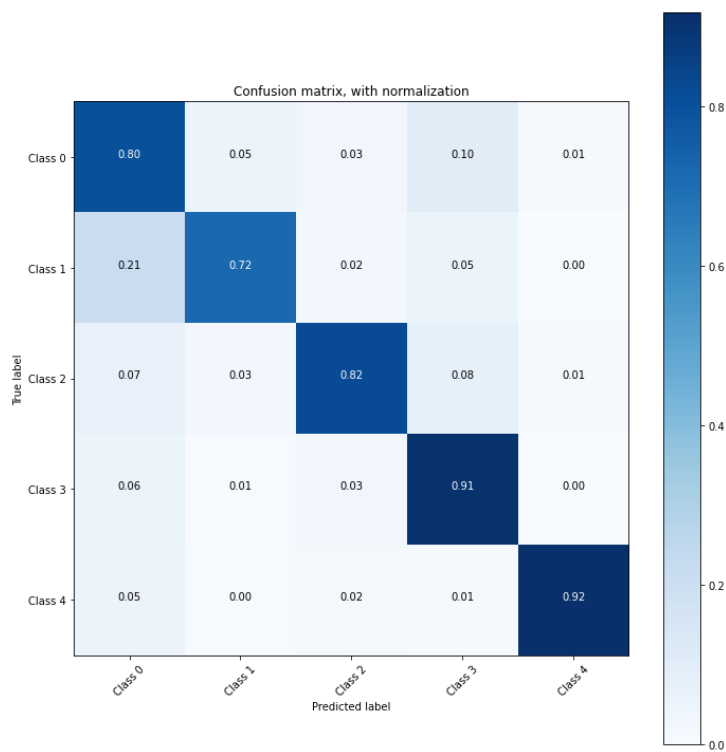

**Figure 6.** Confusion matrix of augmented and balanced MIT-BIH Arrhythmia Dataset

Although the Accuracy values in Table 2 seems high, especially when the Recall and F1-Score values are considered, it seems that the model failed in the classification process. In addition, when the confusion matrix in Figure 5 is examined, it can be observed that the model trained with the unbalanced dataset makes serious errors especially in Class1, Class2 and Class3. Recall and F1-Score values in Table 3 were improved when the dataset was balanced with the new data synthesized with GAN. In addition, this difference is clearly seen from the confusion matrix results in Figure 6. The reason why the accuracy value seems to decrease after the dataset is balanced is because Class0 dominates the model in unbalanced experiment, making the system appear as if it is successful. However, the separation of disease classes in the unbalanced case is unsuccessful.

**CONCLUSION**

As a result of this study, it is shown how the MIT-BIH Arrhythmia Dataset can be transformed from unbalanced to balanced with synthesizing new ECG signals by LSTM-GAN and then how the obtained new balanced dataset can be successfully classified using a 1D CNN architecture. The GAN based data augmentation system designed in this study can be used to create any number of synthetic but realistic data for any ECG arrhythmia studies.